\documentclass[9pt,twocolumn,twoside]{osajnl}
\pdfoutput=1
\journal{ao} 

\setboolean{shortarticle}{false}

\ifthenelse{\boolean{shortarticle}}{\colorlet{color2}{color2b}}{\colorlet{color2}{color2}} 

\dates{Compiled \today}

\ociscodes{(350.1260) Astronomical optics; (350.2450) Filters, absorption}

\doi{\url{http://dx.doi.org/10.1364/ao.XX.XXXXXX}}
\bibpunct{}{}{,}{s}{}{,}

\begin{document} 

\newcommand{\hdblarrow}{H\makebox[0.9ex][l]{$\downdownarrows$}-}

\title{Composite Reflective/Absorptive IR-Blocking Filters Embedded in Metamaterial Antireflection Coated Silicon}

\author[1,*]{C. D. Munson}
\author[2]{S. K. Choi}
\author[1]{K. P. Coughlin}
\author[1]{J. J. McMahon}
\author[3]{K. H. Miller}
\author[2]{L. A. Page}
\author[3]{E. J. Wollack}

\affil[1]{Department of Physics, University of Michigan, 450 Church St., Ann Arbor, MI, 48109}
\affil[2]{Department of Physics, Princeton University, Princeton, NJ, 08544}
\affil[3]{NASA Goddard Space Flight Center, Greenbelt, MD, 20771}

\affil[*]{Corresponding author: cdmunson@umich.edu}

\begin{abstract}

Infrared (IR) blocking filters are crucial for controlling the radiative loading on cryogenic systems and for optimizing the sensitivity of bolometric detectors in the far-IR.   We present a new IR filter approach  based on a combination of patterned frequency selective structures on silicon and a thin (50 $\mu \textrm{m}$ thick) absorptive composite based on powdered reststrahlen absorbing materials. For a 300 K blackbody, this combination reflects $\sim$50\% of the incoming light and blocks \textgreater 99.8\% of the total power with negligible thermal gradients and excellent low frequency transmission. This allows for a reduction in the IR thermal loading to negligible levels in a single cold filter. These composite filters are fabricated on silicon substrates which provide excellent thermal transport laterally through the filter and ensure that the entire area of the absorptive filter stays near the bath temperature. A metamaterial antireflection coating cut into these substrates reduces in-band reflections to below 1\%, and the in-band absorption of the powder mix is below 1\% for signal bands below 750 GHz. This type of filter can be directly incorporated into silicon refractive optical elements.


\end{abstract}

\maketitle

\section{Introduction}

\paragraph*{}
IR blocking filters are critical for optimizing bolometer-based receivers in the millimeter and submillimeter spectral region. In these bands, the IR power emitted from the telescope and surroundings (typically 250-300 K for ground and balloon-based instruments) is much brighter than the sky background. It is therefore crucial to control this out-of-band power. Filters serve to substantially reduce what would be a dominant source of noise and minimize the radiative loading on the cryogenic system. In addition, the filters must not significantly radiate, reflect, or scatter in the band of interest. This requires high thermal conductivity and good heat sinking for any filter containing absorptive components.

\paragraph*{}

Meeting all these requirements in one system is difficult. A variety of approaches have been used previously. Reflective frequency selective surfaces (e.g., patterned onto thin plastic substrates and stacked, as pioneered by Ulrich \cite{Ulrich:filters, Ulrich:grids}, and summarized by Ade \cite{ade:meshfilters}), bulk filters of absorptive materials \cite{Inoue:14, Bock95}, and scattering filters \cite{Manley:powder} are commonly employed. These filtering approaches, however, are not without their complications. The reflective filters patterned on plastic substrates are subject to the intrinsic limits of multi-layered reflectors\cite{multilayer-reflect} as well as heating due to absorption of the plastic. In practice, additional reflective layers improve the filter rejection only incrementally and absorption leads to reradiated power that falls onto detectors\cite{ade:meshfilters}. Absorptive filters, such as those made of bulk Alumina, present difficulties with antireflection coatings, requiring the use of lossy materials that reduce the overall transmission in the desired pass-band by 5\% or more \cite{Inoue:14}. Additionally, Alumina has reststrahlen bands that open up upon cooling, diminishing its overall filtering performance \cite{stierwalt}. In this work we present an example of a hybrid approach based on a combination of  reflective frequency selective structures patterned on silicon substrates, scattering and absorptive layers based on composites of powdered crystals exhibiting the reststrahlen effect, and metamaterial antireflection coatings to control the in-band reflections.  We present a particular example of the construction and performance of a blocking filter designed to pass the 70-170 GHz band in Section \ref{sec:filterconstruction}, and discuss its performance in Section \ref{sec:filterperformance}. We conclude with a discussion of the scalability and applicability of this design in Section \ref{sec:applications}.

\section{Composite Filter Construction}
\label{sec:filterconstruction}

\begin{figure}
	\centering
	\includegraphics[%
 	 width=1.0\linewidth]{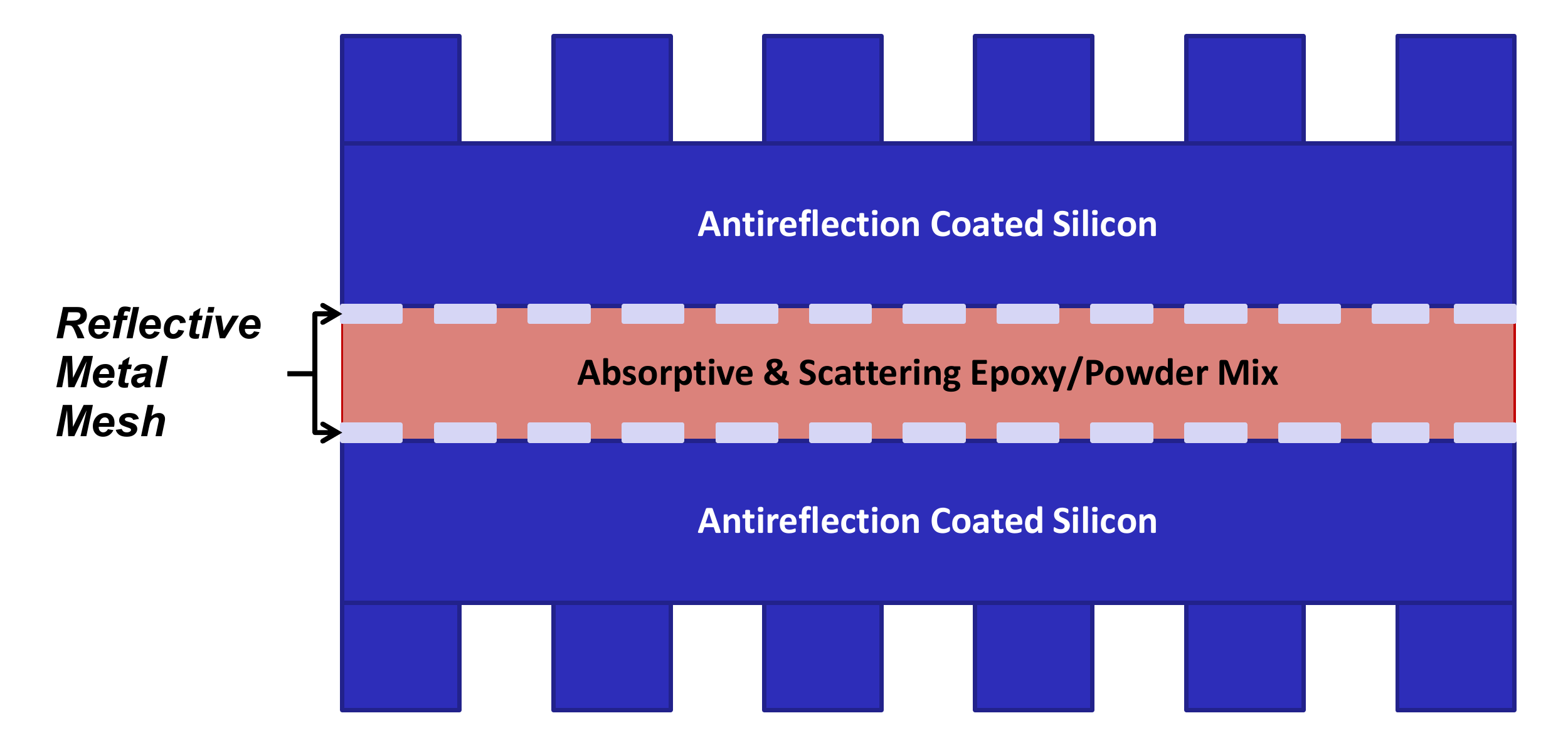}
	\caption{Composite IR-blocking filter construction: The silicon-substrate composite filter is composed of several components. In the pass-band the metamaterial antireflection coated silicon couples light into and out of the filter stack from free space.  In the stop-band, a set of lithographically patterned reflective metal features reflect a significant portion of the incident light, and an absorptive and scattering layer of optical epoxy loaded with powdered reststrahlen materials blocks much of the remaining light.}
\label{filter-schematic}
\end{figure}

Figure \ref{filter-schematic} shows the anatomy of the composite absorptive/reflective IR-blocking filter. This filter consists of  lithographically defined  frequency selective surfaces patterned on two silicon wafers, a 25 $\mu \textrm{m}$ layer of an absorptive mixture of epoxy and reststrahlen powders placed between the two patterned surfaces, and a metamaterial antireflection coating on both vacuum silicon interfaces. At IR wavelengths, light is scattered and reflected off the front silicon wafer and frequency selective surface. The front metamaterial surface scatters light both specularly and diffusely for frequencies above the single-moded limit of the structure. For the metamaterial surfaces described here (tuned to pass 1-2 mm wavelengths), this frequency falls well below the infrared emission from a 300 K blackbody, which peaks at 10 $\mu \textrm{m}$ (30 THz). Infrared light that passes the frequency selective metal mesh is subject to both scattering and absorption by the reststrahlen-epoxy composite. An additional metal mesh layer reflects most of the remaining infrared light back into the epoxy-powder layer, boosting absorption and (to a lesser extent) reflection. This approach reduces the load on the cryogenic stage by reflecting a significant portion of the IR power, and uses an absorbing layer to further attenuate IR power passing the first reflective layer.

At millimeter and submillimeter wavelengths the metal mesh frequency selective surfaces and metamaterial silicon have a high transmission, and the extinction length of the epoxy mixture is inconsequential, leading to low absorption.   Thus in the bands of interest, this structure behaves nearly as if it were a slab of solid low loss silicon treated with a high quality antireflection coating \cite{Datta:13}.

In the remainder of this section we describe the design and performance of the frequency selective surfaces, powder-epoxy mixes, metamaterial antireflection coatings, and conclude with predictions for the integrated performance of our filters.

\subsection{Frequency Selective Metal Mesh Filters}
\label{sec:MMF}

\begin{figure}
\begin{center}
\includegraphics[%
  width=1.0\linewidth,
  keepaspectratio]{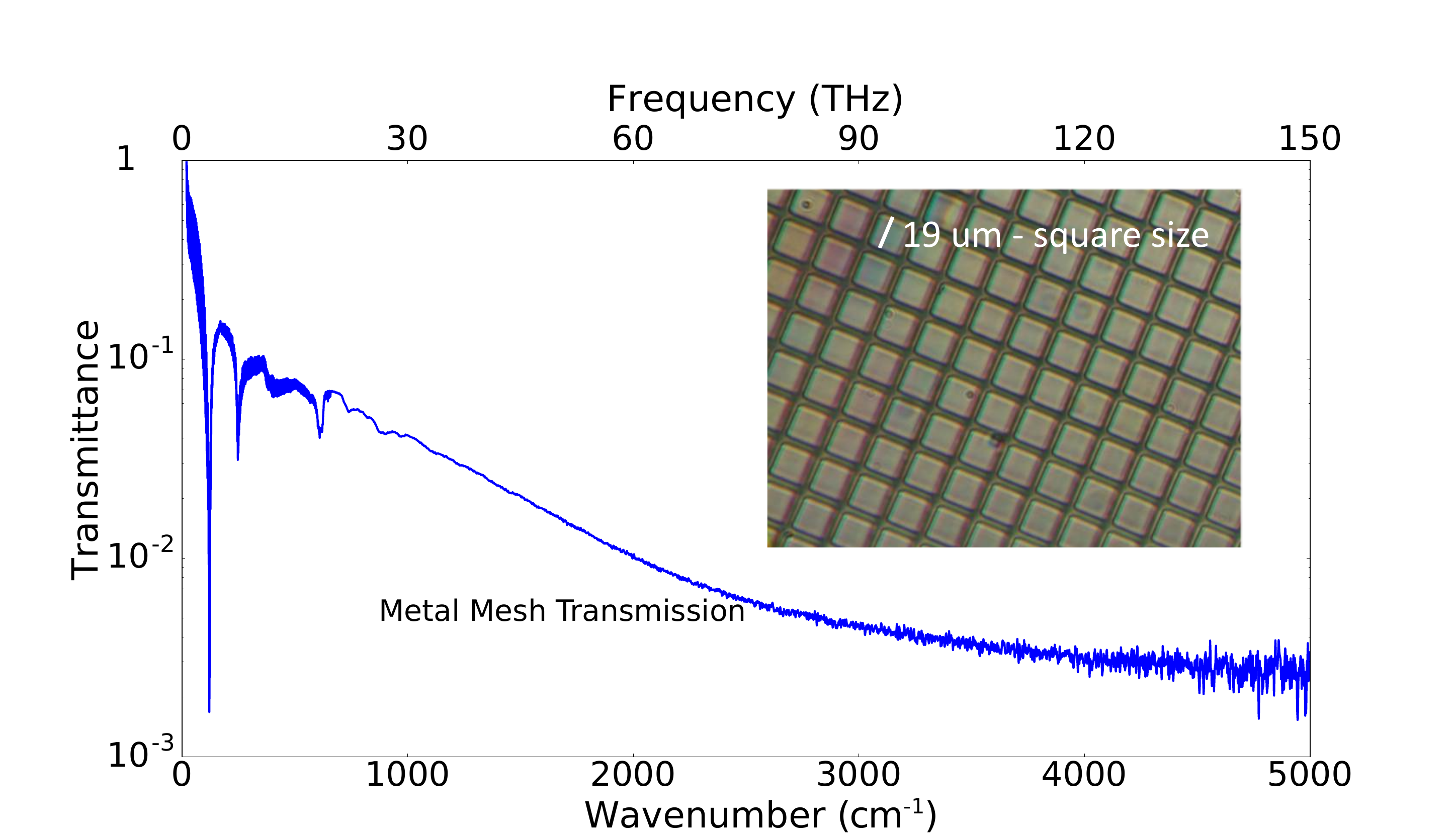}
\end{center}
\caption{The measured reflectance of a frequency selective surface formed by lithographically patterned aluminum squares on a silicon substrate. Note the null at 120 cm$^{-1}$, corresponding to the target cutoff frequency of the filter. The filling fraction is 63\%, with additional high-frequency reflectivity resulting from the silicon substrate. Inset, a photograph of this sample.}
\label{Mesh-picture}
\end{figure}

The first filtering component of these composite IR-blocking filters is a low-pass frequency selective surface formed by a mesh of resonant metal squares. These squares act as a grid of capacitive elements and pass low frequencies while reflecting high frequencies. In the low frequency limit, the metallization layer is effectively nonexistent giving nearly unity transmission, and in the high frequency limit, these features reflect according to the fill factor of the metallization. In the resonant region between, there is some additional reflection, with the cutoff frequency set by the grid spacing. We selected the grid parameters to place the cutoff frequency well above the upper end of our desired signal band (170 GHz), but below the peak emission frequency of a 300 K blackbody (18 THz). We selected a grid period of 23.8 $\mu \textrm{m}$ and street widths of 5 $\mu \textrm{m}$, for a cutoff frequency in silicon of $\sim$3.6 THz (corresponding to a freespace wavenumber of 120 cm$^{-1}$ \cite{Ulrich:grids}, clearly visible as the first resonance in the measurement in Figure \ref{Mesh-picture}). 
These dimensions were additionally constrained to be within the capabilities of large-diameter liftoff lithographic techniques (limiting us to minimum features of few micron scales). A prototype of this design was thoroughly characterized (see Figure \ref{Mesh-picture}), and the performance is in good agreement with our theoretical expectations, with a total reflectivity of  83\% for a 300 K blackbody. 

These features are straightforward to tune to a desired cutoff frequency and can be fabricated on the silicon using standard lithographic techniques. Design of these features was carried out via the analytical techniques described by Ulrich \cite{Ulrich:filters}, with additional optimization and verification via modeling in ANSYS HFSS \cite{Ansys:HFSS}.

\subsection{Reststrahlen Materials and Powder Filters}

\paragraph*{}

\begin{figure*}[!ht]
\begin{center}
\includegraphics[%
  width=1.0\textwidth, keepaspectratio]{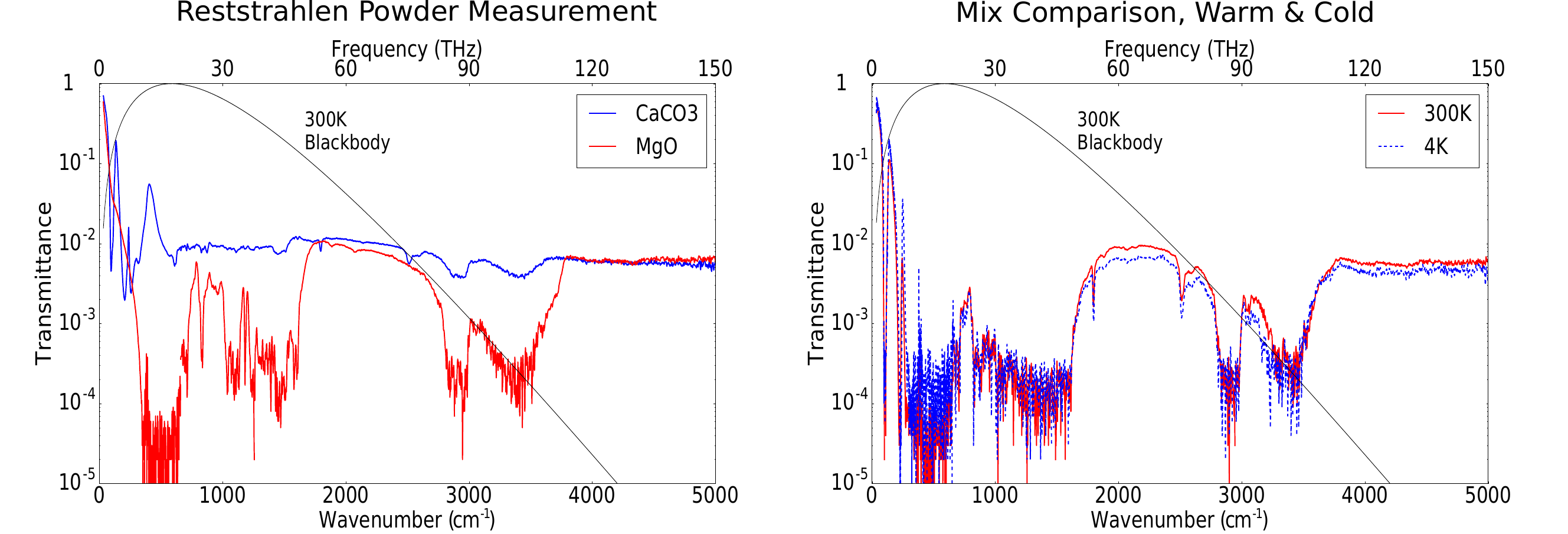}
\end{center}
\caption{Left: An FTS measurement of $CaCO_3$ and $MgO$ powder filters, demonstrating good blocking performance and reproduction of the characteristic reststrahlen absorption bands. Right: The measured performance of a $CaCO_3$/$MgO$ mix powder filter warm (at 300 K) and cold (at 4 K) showing minimal changes to the absorption characteristics and demonstrating suitability of this technology for cryogenic IR blocking filters. }
\label{reststrahlen}
\end{figure*}

The reststrahlen effect, from the German for ``residual rays," is the absorption of light at characteristic frequencies \cite{Fermi:mcqs} of the bound ion pairs in crystalline materials that falls in the infrared \cite{YAMADA:62, Robinson:FIR}. These structural resonances prevent light in the reststrahlen band from propagating through the material, and effective low-pass filters have been previously demonstrated using powdered reststrahlen materials \cite{YAMADA:62}. Transmissive and reflective filters have also been realized from bulk reststrahlen crystals \cite{Robinson:FIR}.

\paragraph*{}

Powder filters, formed by reststrahlen powders mixed into a polyethylene carrier, were demonstrated as IR blocking filters by Yamada et al. \cite{YAMADA:62} These filters, due to their plastic carrier, suffered from heating and reradiation. We fabricated similar powder filters by mixing reststrahlen powders into a mixture of toluene thinned Epotek 301 optical epoxy and applying thin layers with a commercial spraygun mounted on a robotic gantry. This allowed for the creation of thin (~25 $\mu \textrm{m}$) uniform layers of the epoxy-powder composite. The powders were chosen from the assortment of materials characterized by Yamada \cite{YAMADA:62} to provide good blocking coverage for wavelengths from \textless 10 $\mu \textrm{m}$ up to 150 $\mu \textrm{m}$ (corresponding to 67-1000+ cm$^{-1}$, 2-30+ THz), overlapping transmissive regions of one powder with absorptive regions of another. Additionally, the powders were chosen for ease of use and acquisition, limiting us to non-hazardous materials (excluding materials such as thallium and beryllium salts, which have been previously used \cite{YAMADA:62}) that are common chemical reagents. As a result of these constraints, our filters consist of magnesium oxide ($MgO$) and calcium carbonate ($CaCO_3$), both with 5-20 $\mu \textrm{m}$ typical particle sizes. See Table \ref{table:1} for the composition of the reststrahlen composite layer, obtained from the Maxwell-Garnett theory \citep{EffectiveMedium}. The measured optical transmissions for these powder filters are shown in Figure \ref{reststrahlen}. These transmission spectra exhibit the characteristic absorption features expected for both materials \cite{YAMADA:62, Robinson:FIR}. We combined these powders in equal parts (by mass) to toluene thinned optical epoxy and applied it with the spraygun. For the full composite filter, this epoxy layer also adhered the two silicon wafers together. The particle size is sufficiently small that the epoxy layer can be treated as a dielectric mixture that is well described by mean field effective medium approximations \cite{EffectiveMedium} in the instrument band of 70-170 GHz (see Figure \ref{LF}).

\begin{table}[!h]
\centering
\begin{tabular}{ |c|c|c|c| }
\hline
Material & Particle Size & Volume Fraction & $\epsilon_r$ \\
\hline
$CaCO_3$ & 5-20 $\mu \textrm{m}$ & 0.092 & 8.45 \\
$MgO$ & 5-20 $\mu \textrm{m}$ & 0.073 & 9.8 \\
Epotek 301 & N/A & 0.835 & 3.7 + 0.1i \\
Mixture & N/A & 1.00 & 4.35 + 0.095i \\
\hline
\end{tabular}
\caption{The composition of the epoxy composite layer.}
\label{table:1}

\end{table}

\paragraph*{}
The performance was characterized at cryogenic temperatures to ensure their proper cryogenic functioning. It is known that some reststrahlen materials have absorption bands that open up when the material is cooled down. In particular, alumina ($Al_2 O_3$) is known to have a section of its absorption band (between 30 and 300 $\mu \textrm{m}$) open up at temperatures of tens of Kelvin \cite{Dobrov:sapphire, Hadni:65}. To explore whether this would be problematic with our materials, a powder filter consisting of our mixture of $CaCO_3$ and $MgO$ was measured in a Fourier Transform Spectrometer (FTS) at a range of temperatures between 4 K and 300 K. Plots of the extremal temperatures are shown in the right panel of Figure \ref{reststrahlen}. From this, it is apparent that changes in the absorption spectrum are minimal, and do not compromise the filtering performance of a 300 K blackbody significantly.

\subsection{Metamaterial Antireflection Coated Silicon}

\begin{figure}
\begin{center}
\includegraphics[%
  width=1.0\linewidth,
  keepaspectratio]{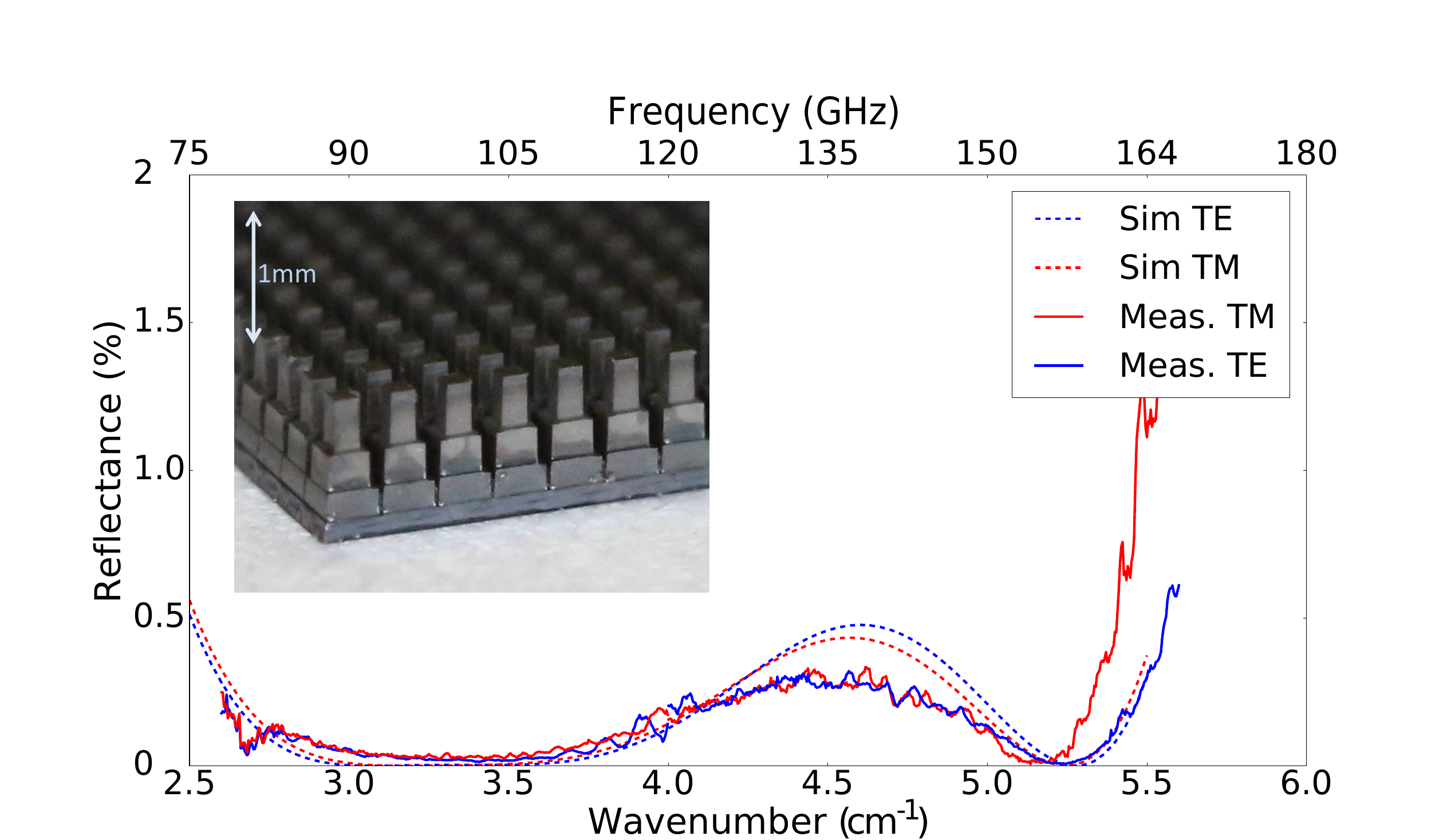}
\end{center}
\caption{Metamaterial antireflection coated silicon: The performance of a three-layer metamaterial surface for the 90 GHz and 150 GHz Cosmic Microwave Background observing bands, demonstrating good agreement with simulated performance (from HFSS) and reflections below 0.5\% averaged across the signal band. The simulation and measurement are made for a 15$^{\circ}$ angle of incidence. Inset: A photograph of a sample of metamaterial antireflection coated silicon, consisting of three simulated dielectric layers of machined sub-wavelength features.}
\label{AR-picture}
\end{figure}

High resistivity silicon is an excellent material for millimeter and submillimeter wave optics. The low loss (tan $\delta$ \textless 7e-5) \cite{Datta:13} and high refractive index allows for refractive optics with negligible loss. High purity, single-crystal silicon is available in large diameters and can be readily obtained. Additionally, silicon has a high thermal conductivity ($>2\,\rm kW / m  K$)\cite{glassbrenner:si}, which prevents filters with an absorptive component from heating up. 
The high refractive index of silicon presents the problem of high reflectivity for optics, but this problem has been successfully managed with a machined sub-wavelength anti-reflective surface on the outer surfaces of the optics. These features allow for the creation of antireflection ``coatings" with simulated dielectric layers. These metamaterial antirreflection surfaces allow for high transmission optics (\textgreater 99\% transmission across a 70-170 GHz band for a three-layer simulated dielectric coating \cite{Datta:ACTPol}) fabricated entirely from silicon. A thorough discussion of this antireflection surface treatment approach is given in Datta et al \cite{Datta:13}. Figure \ref{AR-picture} shows a photograph of a three-layer "coating" as well as a comparison of the simulated and measured reflection at 15$^{\circ}$ incidence for two orthogonal polarizations.

\section{Composite Filter Performance}
\label{sec:filterperformance}

The spectral performance of our composite filters  and their constituent components was evaluated using FTS measurements. The thermal performance and integrated measurements were made in a cryostat open to a 300 K blackbody, using a carbon loaded disk bolometer. The performance of the individual components was measured across both the low-frequency signal band (down to 10 cm$^{-1}$, 300 GHz) limited by the low frequency capability of the FTS) and high-frequency blocking band (up to 5000 cm$^{-1}$, 150 THz). Additionally, the overall performance of a full composite was measured across the blocking band. 

\paragraph{Measurement Methods:}

The reflectance and transmittance in the range 10-5,000 cm$^{-1}$ (0.3-150 THz) were measured with the Bruker IFS 125 FTS using the following two combinations of source-beamsplitter-detector: Hg arc lamp—multilayer mylar—liquid helium cooled bolometer (30-700 cm$^{-1}$, 0.9-21 THz), and globar—Ge-coated KBr—DLATGS (500-5,000  cm$^{-1}$, 150-1500 THz).  The spectral response in the overlap region agreed to within 0.5\%.  The data sets were merged into one spectra by equating the areas underneath the curves in the overlap region using a weighted average.  The reflectance was measured in a collimated beam geometry with an $8^{\circ}°$ angle of incidence.  The transmittance was measured in a focused beam (f/6.5) geometry at normal incidence.  The total hemispherical transmittance in the range 500-5,000 cm$^{-1}$ was collected using a Bruker-made integrating sphere (75 mm diameter) accessory for the FTS with its own internal DLATGS detector.  A diffusely reflecting gold surface, which matches the inner surface of the sphere, was placed over the sample port to collect the reference scan.

For integrated testing, a cryostat was used to measure the radiative properties and total infrared blocking.

\begin{figure*}
\begin{center}
\includegraphics[%
  width=0.96\textwidth,
  keepaspectratio]{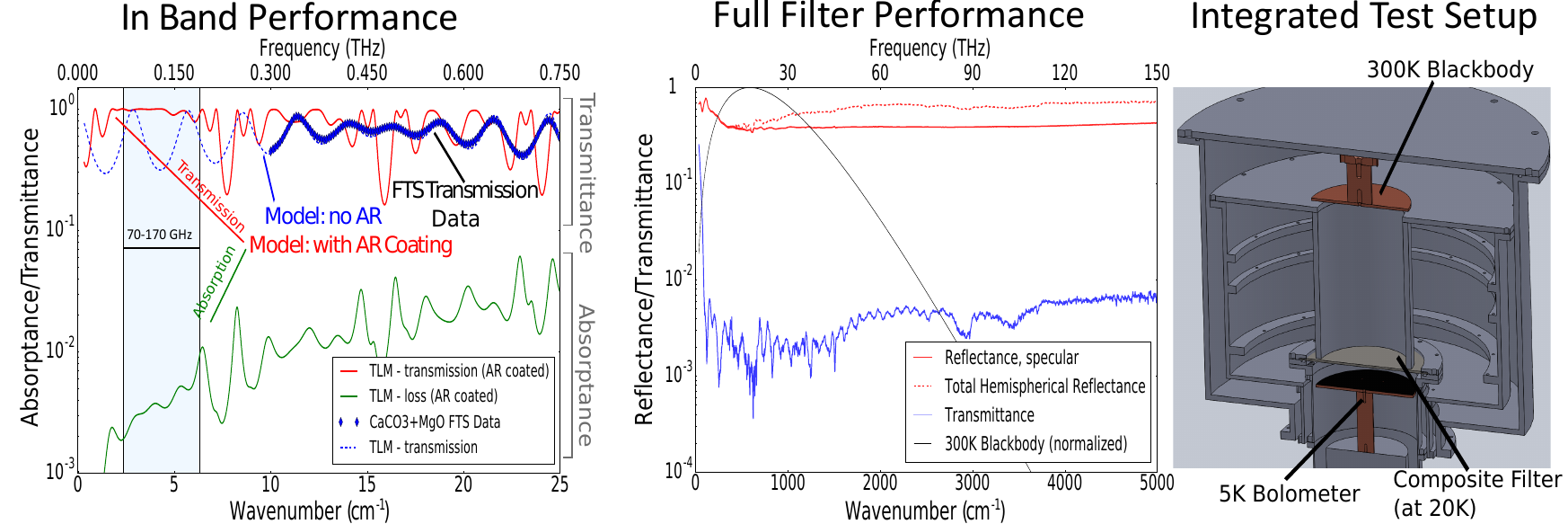}
\end{center}
\caption{The performance of composite filter components. Left: The measured (blue diamonds), and simulated (blue dashed) low frequency performance for a single 75 $\mu \textrm{m}$ layer of reststrahlen powder mix, in epoxy, on a silicon wafer. Additionally shown are the best fit simulated performance (transmission and absorption) for a stack consisting of AR coated silicon on either side of a 75 $\mu \textrm{m}$ powder mix layer. The target transmission band for the AR coating is 70-170 GHz and is marked with the blue band on the plot. Middle: The IR blocking performance of a full composite filter is shown, with a 300 K blackbody overlaid. Right: A drawing of the integrated test cryostat, wherein the filter (held at 20 K) is used to block power from a 300 K blackbody falling on a 5 K disk bolometer (black). A total blocking efficiency of \textgreater 98\% was demonstrated using this setup, agreeing with the value derived from the FTS measurements. This was determined by measuring the heating of the bolometer when exposed to an aperture open to 300 K, and determining the incident power relative to the same environment without a blocking filter. }
\label{LF}
\end{figure*}

\paragraph{IR Blocking Performance:}

The infrared blocking performance of these filters was measured on the FTS up to 5000 cm$^{-1}$ (2 $\mu \textrm{m}$ wavelength, 150 THz), giving a full characterization of the transmission across the spectrum of a 300 K blackbody.  These measurements reproduce the characteristic reststrahlen powder filter shape in thicker (75 $\mu \textrm{m}$) layers \cite{YAMADA:62} (see Figure \ref{reststrahlen}), and demonstrate excellent IR blocking in layers as thin as 25 $\mu \textrm{m}$. 

In the FTS measurements, the composite filter specularly reflected \textgreater 40\% of the light incident from a 300 K blackbody (indicating reflection off the front silicon surface and metal mesh features), and diffusely reflected another ~10\%, indicative of backscattering off the powder layer. The overall reflectance was lower than would be naively expected from the single reflector measurements in Section \ref{sec:filterconstruction}. This is likely explained by coupling of fields from the reflectors into the lossy epoxy composite below. Should higher reflectivity be needed, the reflective layers could be separated from the lossy composite and multiple reflective layers could be employed.

\paragraph{Low Frequency Performance:}

The low frequency performance (below 1 THz) of a 75 $\mu \textrm{m}$ layer of the powder filter component was measured down to 10 cm$^{-1}$ (300 GHz) using an FTS. These data were then fit with a simple transmission line model. The index of the powder layer was estimated via a Maxwell-Garnett effective medium approximation \cite{Choy:effmed}, and treated as a layer of continuous material of an unknown thickness. The thickness of this layer and the underlying silicon layer were then fit using a standard least-squares approach. For a 75 $\mu \textrm{m}$ mix layer (consisting of a binder of Epotek 301, $\epsilon_r = 3.7 + 0.1i$ as measured in the 1-10 THz range with a sample etalon formed by silicon wafers, loaded with powdered $MgO$, $\epsilon_r = 9.8$, and $CaCO_3$, $\epsilon_r = 8.45$ in 0.073 and 0.092 volumetric fractions respectively) on 500 $\mu m$ thick silicon ($\epsilon_r = 11.67$), the model accurately reproduced the measured total thickness. This model was then used to extrapolate to lower frequencies and simulate the effect of adding a three-layer antireflective coating to a filter made using this material. This model shows that the reststrahlen powder filter introduces minimal loss (dominated by the epoxy carrier) in a signal band from 70-170 GHz, and that an instrument-band transmission of \textgreater 99\% should be achievable for a filter using this technology (with the total transmission limited by the antireflection coating performance). In the low frequency region, the filter performance is well represented with a simple transmission line model, taking into account the effective index of the composite determined from an effective medium approximation. The measured and modeled low frequency performance is shown in Figure \ref{LF}.

\paragraph{Thermal Performance and Cryostat testing:}

An integrated test of the composite filter performance was carried out in a 3-stage cryostat, to measure the total blocking efficiency of a 15 cm diameter prototype (Figure \ref{LF}). A 5.8 cm diameter disk bolometer (made from carbon-blackened copper) was held at a base temperature of 5 K. The filter was mounted on the 20 K stage directly above the bolometer, and blocked light being emitted from a blackbody at 300 K visible through a hole in the lid of the 100 K stage. The total power blocked by the filter was then estimated by measuring the bolometer temperature with and without the filter. Additionally, the heating of the 20 K stage was measured to estimate the power absorbed by the filter, and the filter temperature was measured at its center and edge to characterize the thermal gradient across it. Figure \ref{LF} shows a schematic of the cryostat. 

There was negligible heating of the center of the filter (with thermal gradients of less than 1 K from center to edge) when the filter was mounted to the 20 K stage and used to block the power from a 7 cm diameter window open to 300 K. The filter as a whole heated to 2 K above the stage temperature, demonstrating effective heat-sinking and power removal from the filter as well. In this configuration, the power deposited on a carbon disk bolometer at 5 K was measured, and this measurement established a lower limit on the blocking of \textgreater 98\% of a 300 K blackbody. This lower limit is in agreement with the FTS measurements of the full composite filter.

\paragraph{Comparison to Competing Filter Technologies:}

Compared to existing reflective filters, built of reflective frequency selective surfaces patterned onto plastic substrates and stacked, our filters offer several distinct advantages. The high thermal conductivity of silicon and low absorption means that there is negligible heating of the filter center relative to the heatsunk edge. Additionally, silicon is a stiffer, more mechanically robust substrate, which allows for better control of lithographic features and finished filters that are stronger and less susceptible to deformation and wrinkling due to thermal cycling. The composite construction of our filters additionally integrates absorptive components to improve the overall out-of-band rejection beyond what can be attained with a reasonable number of stacked reflective components.

Compared to bulk absorptive filters such as alumina, our filters offer the ability to reflectively reject a significant portion of the power, reducing the load on the cryogenic system. The reststrahlen materials we used are also easy to obtain. The use of multiple reststrahlen materials allows for selection of complementary sets of materials that improves the frequency coverage and prevents sections of the band from becoming transmissive (such as happens with alumina when cooled). Silicon is a lower loss material than alumina and the antireflection coatings, being metamaterial silicon, are significantly lower loss than the epoxies used to antireflection coat bulk alumina filters, and the thermal conductivity of silicon is significantly higher than alumina or PTFE. Finally, the composite filter offers a number of adjustable parameters not present in bulk filters that allow better tailoring of the filter characteristics to achieve the desired performance. 

\section{Frequency Range of Applicability}
\label{sec:freqscale}

The components of this style of silicon substrate composite filters all have applicability across a broad range of frequencies. The limiting frequencies for the different components vary, but filters can be constructed using these techniques and some subset of the components for frequencies ranging from tens of gigahertz to hundreds of terahertz depending on the components used. The frequency scaling of each of the four important components is discussed in this section. The combination of several of these components can form effective filters across a wide range of frequencies.

\subsection{Frequency Selective Metal Features}
The frequency selective metal mesh features can be fabricated across the full range of sizes available with modern lithographic techniques. This results in applicability across the full range over which silicon is transparent, up to the beginning of the absorption band at 1 $\mu \textrm{m}$ wavelength (300 THz). Silicon offers better control and repeatability of lithographic features than plastic substrates, allowing smaller features (and therefore higher frequencies) to be attained. Additionally, the metal features can be formed with both high- and low-pass frequency responses, as well as with anisotropy, which enable a range of filtering characteristics, including band-defining filters, and filters with polarization dependence.

\subsection{Silicon Substrate}
The high-resistivity silicon substrate offers excellent transmission performance from low frequencies up to the beginning of its absorption band in the near-infrared. It begins to become absorptive at a wavelength of approximately 1 $\mu \textrm{m}$, which corresponds to the plasma frequency (300 THz) in the media\cite{palik-silicon}. At frequencies lower than this (longer wavelengths), it remains highly transmissive and low loss, with typical loss tangent below 0.0002 across the THz region \cite{thzsilicon}, and remaining low through the IR (with the exception of two lattice absorption features at 600 and 650 cm$^{-1}$, corresponding to 18 and 19 THz or wavelengths of 15 and 16 $\mu \textrm{m}$)\cite{mmsilicon} .

\subsection{Antireflective Coatings}
The metamaterial antireflective coatings have applicability to a similarly broad range of frequencies. At lower frequencies, the coatings can be machined into the surfaces using a dicing saw or conventional grinding techniques. This approach to the antireflection coatings has been successfully demonstrated for frequencies up to 1.5 THz with excellent performance.

With increasing frequency, higher precision fabrication approaches become necessary. Using laser machining to cut smaller features is one option, up to the limits of current laser machining capabilities. Additionally, lithographic techniques involving the patterning of the surface and etching of features, such as deep reactive ion etching (DRIE) offer a solution to still higher frequencies, with attainable feature sizes of tens of nanometers that would be suitable for antireflection coatings well above the 1 $\mu \textrm{m}$ absorption cutoff of silicon. 

For higher frequency and non-cryogenic applications, conventional antireflection coating approaches can also be used, substituting layers of bulk dielectric materials (or applied thin films of dielectric materials) for the simulated dielectric formed by the metamaterial silicon. 

\subsection{Scattering and Absorptive Powder}
The scattering and absorptive powder layer is useful only for lower frequencies due to the fixed reststrahlen bands and limits on attainable powder size. The particulate size of the powders can be selected to move the scattering peak to higher or lower frequencies as needed, but the absorption bands of the materials are fixed by the material choice. For some applications, suitable materials are available that will enhance the overall light rejection of the filter. For filters where blocking in these bands is not desired, the scattering and absorptive powder can be removed from the design. The epoxy binder can likewise be removed in favor of direct bonding the wafers, for applications where the epoxy would increase the loss and is not needed as a carrier for scattering and absorptive powders.

\section{Potential Applications}
\label{sec:applications}

In addition to forming effective free-space IR blocking filters, this filtering approach offers several novel possibilities for silicon-substrate optical elements. Lower frequency-selective metal elements can be incorporated into these filters to aid in defining the instrument signal band. Anisotropic application of these filtering techniques can form birefringent materials. These filters can be easily and inexpensively integrated into other optical components, such as silicon lenses. 

Filters with higher cutoff frequencies and better uniformity can also be constructed for effectively blocking higher frequencies (into the mid-IR and higher) due to the high quality of silicon substrates available (superior material properties and surface finish allow for finer lithography, leading to better high frequency performance, a current limit of plastic-substrate filters \cite{ade:meshfilters}). 
The frequency range can be tuned by adjusting the reflective grid parameters, the scattering particle size, and the specific material used for scatterers, and multi-layer reflective structures can be formed to increase the overall reflectivity. Direct wafer bonding can reduce or eliminate the need for lossy epoxy components, and reflective layers can be precisely spaced using standard lithographic techniques to set the layer thicknesses. Collectively, these techniques will further improve control of the filter properties, enabling higher performance and better customization of the filter characteristics to the desired application.

Finally, more complex metamaterial behaviors can be added through more complicated lithographic features and machined subwavelength features. This enables a wide array of novel optical characteristics with potentially broad consequences for future imaging systems.

\section{Acknowledgments}

This work was supported by a NASA Office of the Chief Technologist’s Space Technology Research Fellowship \# NNX12AM32H. Lithography was performed using the Lurie Nanofabrication Facility at the University of Michigan. JJM was supported by DE-SC0015799 for this work.



\end{document}